\documentclass[aps,preprint,showpacs]{revtex4}
\usepackage{amsfonts}
\usepackage{amsmath}
\usepackage{amssymb}
\usepackage{graphicx}
\begin{document}
\title{Fusion Hindrance in the Heavy Ion Reactions \\ -- Border Between the Normal and Hindered Fusions}
\author{Caiwan Shen$^1$, David Boilley$^{2,3}$, Qingfeng Li$^1$, Junjie Shen$^{1,4}$, Yasuhisa Abe$^5$}
\affiliation{
$^1$School of Science, Huzhou Teachers College, Huzhou 313000, P.R. China \\
$^2$GANIL, CEA/DSM-CNRS/IN2P3, BP 55027, F-14076 Caen cedex 5, France \\
$^3$Univ. Caen, Esplanade de la Paix, B.P. 5186, F-14032 Caen Cedex, France  \\
$^4$School of Science and Information Engineering, Zhejiang Normal University, Jinhua 321000, P.R. China \\
$^5$Research Center for Nuclear Physics, Osaka University, Osaka 567-0047, Japan 
}


\begin{abstract}
The fusion hindrance in heavy ion collisions is studied in the framework 
of the two-center liquid drop model. It appears that the neck and the radial 
degrees of freedom might both be hampered by an inner potential barrier on 
their path between the contact configuration to the compound nucleus. 
Heavy ion reactions with and without the two kinds of fusion hindrance are
classified through systematic calculations. It is found that the number of 
reactions without radial fusion hindrance is much smaller than that
without neck fusion hindrance, and for both kinds of fusion 
hindrance the number of reactions without fusion hindrance at small 
mass-asymmetry parameter $\alpha$ is smaller than that at large $\alpha$.
In the formation of a given compound nucleus, if a reaction with $\alpha_c$ 
is not hindered, then other reactions with $\alpha > \alpha_c$ are also not
hindered as it is well known experimentally.

\end{abstract}
\maketitle

\section{ Introduction}

The fusion hindrance that appears in heavy ion reactions has been known for many
years. Now the mechanism of fusion hindrance is gradually understood as to be due to an
extra internal barrier between the touching configuration and the compound shape after
overcoming the Coulomb barrier \cite{bjornholm,royer,blocki}. However what
causes the internal barrier is a difficult problem. One of the hypotheses is
that the internal barrier could be thought as the conditional saddle point in
the liquid-drop potential as well as could be attributed to an effective
barrier due to the dissipation of the incident kinetic energy
\cite{gross,frobrich}. 

Since there are two barriers for the fusion, its theoretical description 
is divided into two consecutive steps: one is from infinity to the contact 
configuration passing over the Coulomb barrier, and another one is from the
contact point to the compound state overcoming the internal barrier. This 
so-called two-step model explains the extra-push energy and furthermore 
gives an energy dependent fusion cross
sections \cite{abe1,shen1,abe2,shen2}. The measured residue cross section 
can be calculated by plugging a statistical evaporation model on the two-step model.

To evaluate the amplitude of the hindrance to the fusion, the size of 
the inner barrier is not enough. The dissipation mechanism also plays an 
important role and one has to solve dynamically the diffusion over this 
barrier. This was done analytically in one dimensional case assuming a 
parabolic barrier \cite{abe3, boilley0}. Here, we will focus our attention 
on the appearance of the hindrance to the fusion that is due to the saddle 
point height measured from the energy of the touching configuration formed 
by the projectile and target nuclei of the incident channel. In other words, 
there is no hindrance for cases without internal barrier. Therefore, the 
border between the normal and the hindered fusion is given by the condition 
that the contact configuration be on the top of the saddle point.

For mass-symmetric reactions, the fusion path between the touching configuration 
and the compound shape can be easily determined since the system evolves only 
along the radial degree of freedom. In order to determine if the fusion hindrance
occurred for a given reaction, we simply compare the position of the 
barrier and the touching configuration: the fusion exists while the barrier 
is located inner side of the touching point and vice versa. Ref. \cite{shen3} 
systematically analyzed different symmetric reactions and drew a clear border 
between the two kinds of reactions (with and without fusion hindrance).

How is the situation for mass-asymmetric cases? Generally this problem becomes 
difficult because the system crosses a ridgeline, not a saddle. It is then 
difficult to determine an effective reaction coordinate. However there is a 
way to avoid the difficulty: let the reactions evolve starting from the contact
point on the potential surface without any random force, and then check if the 
compound state is reached. If the composite system evolves toward the compound 
direction, it means that there is no barrier between contact point and compound 
state, and thus the fusion hindrance does not exist. Otherwise, the fusion 
hindrance occurs. With this method, the current paper studies the occurrence 
of the fusion hindrance for asymmetric reactions.

Refs. \cite{abeneck,boilley1,boilley2} show that the neck of the amalgamated system 
also plays an important role in the hindrance to the fusion. Thus, one of 
the related questions is about the evolution of neck degree of freedom: 
Is there also a fusion hindrance in the neck evolution, like for radial evolution? 

We will answer to these questions in the present paper.

\section{Parameterization of the composite system}

There are several ways to parameterize the shape of the amalgamated system. 
In this paper we use the so-called two-center parameterization \cite{sato,maruhn}
based on three important parameters which are: the distance between two 
centers $z$, the mass-asymmetry parameter $\alpha$, and the neck parameter 
$\varepsilon$. The dimensionless parameter $z$ is defined as follows:
\begin{equation}
z=R/R_{0},
\label{eq-1}
\end{equation}
where $R$ denotes the distance between two centers of the harmonic potentials, 
and $R_{0}$ the radius of the spherical compound nucleus. The mass-asymmetry 
parameter is defined as usual,
\begin{equation}
\alpha=\frac{A_{1}-A_{2}}{A_{1}+A_{2}},
\label{eq-2}
\end{equation}
where $A_{1}$ and $A_{2}$ are the mass numbers of the colliding nuclei. 
The neck parameter $\varepsilon$ is defined by the ratio of the smoothed
height at the connection point of the two harmonic potentials and that 
of spiked potential. In this description, nuclear shape is defined by 
equipotential surfaces with a constant volume. For example, $\varepsilon=1$ 
means no correction, i.e., complete di-nucleus shape, while $\varepsilon=0$ 
means no spike, i.e., flatly connected potential, which describes highly 
deformed mono-nucleus. Thus, the neck describes shape evolution of the compound
system from di-nucleus to mono-nucleus. The initial value of $z$, $\alpha$ and 
$\varepsilon$, for a given reaction at the touching point is
\begin{align}
z_{0} &  =\frac{A_{\rm{p}}^{1/3}+A_{\rm{t}}^{1/3}}{(A_{\rm{p}}
+A_{\rm{t}})^{1/3}},\label{eq-2.4}\\
\alpha_{0} &  =\frac{A_{\rm{t}}-A_{\rm{p}}}{A_{\rm{t}
}+A_{\rm{p}}},\label{eq-2.5}\\
\varepsilon_{0} &  =1,\label{eq-2.6}
\end{align}
respectively. Here $A_{\rm{p}}$ ($A_{\rm{t}}$) stands for the mass
number of the projectile (target). With this parameterization, the liquid drop 
model (LDM) potential is adopted to calculate the energy of the touching 
di-nucleus.

These three parameters are connected by the potential, the 
inertia and friction tensors. For the sake of comprehension, we will study 
the neck and the radial degrees of freedom separately.

\section{Fusion hindrance in the neck evolution}

In this section we focus our attention on the potential landscape seen by 
the neck parameter. Let us recall that in the two-step model, the fusion 
hindrance is described as the additional barrier hampering the evolution 
from the contact point to the compound nucleus.

\begin{figure}[ptbh]
\begin{center}
\includegraphics[height=2.7215in,width=2.8659in]{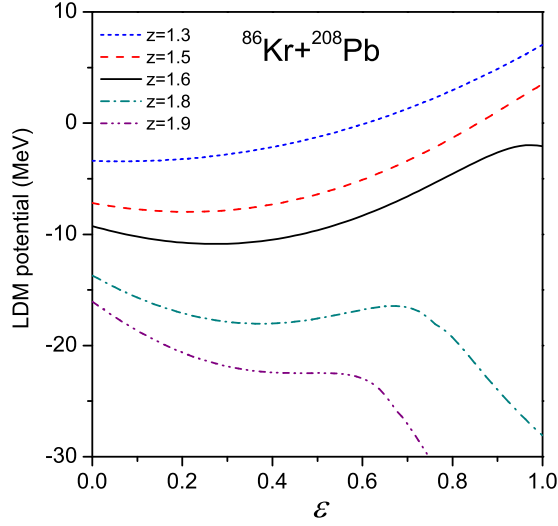}
\caption{LDM potential as a function of the neck parameter $\varepsilon$ 
for $^{86}$Kr+$^{208}$Pb at various relative distances.}
\label{fig-1}
\end{center}
\end{figure}

In order to show the direction of evolution of the neck degree of freedom, 
the LDM potential is plotted as a function of $\varepsilon$ in 
Fig. \ref{fig-1}, where the different lines correspond to different 
distances between the projectile and target. The figure shows two 
different directions of evolution: (i) at small $z$ ($z = 1.3$ or $z = 1.5$) 
the potential decreases with reducing neck parameter (thicker neck) 
which means that at contact the neck becomes thick by the driving effect
of the potential. (ii) at large $z$ ($z =1.8$ or $z = 1.9$), the potential 
increases with reducing neck parameter around $\varepsilon\sim1$ and a 
bump exists at a critical value of $\varepsilon$. This indicates that an 
extra energy should be provided to reach the thick neck, i.e. to form a 
mono-nucleus. In other words, the system is hindered to form a compound 
nucleus. In fact, the frontier between the two cases can be determined 
by the slope of the potential at the point $\varepsilon = 1$:
\begin{equation}
f=\left.  \frac{dV}{d\varepsilon}\right\vert _{\varepsilon=1}.
\label{eq-3}
\end{equation}
Here $V$ denotes for the LDM potential. Positive $f$ and negative $f$ 
correspond to the first case and the second case, respectively.

\begin{figure}[ptb]
\begin{center}
\includegraphics[height=2.2125in,width=6.3628in]{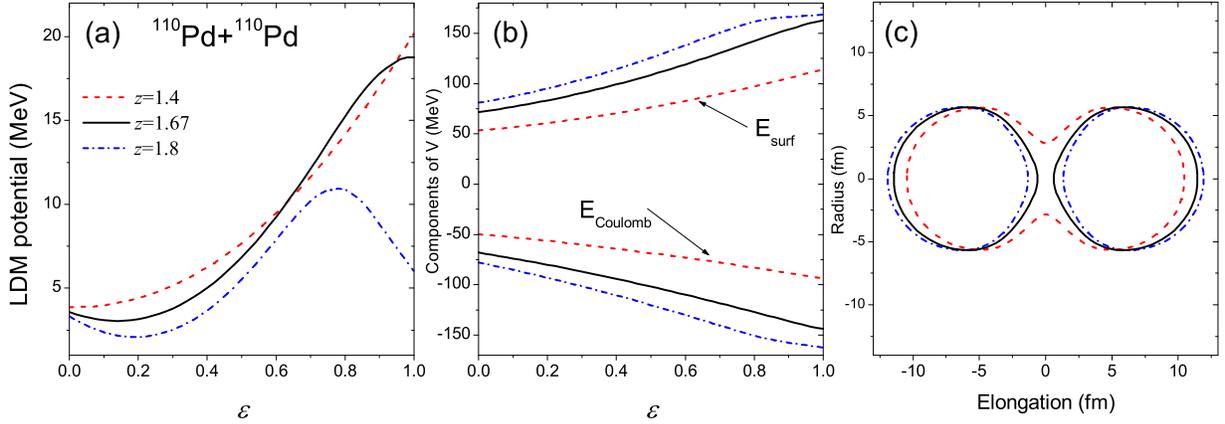}
\caption{LDM potential for $^{110}$Pd + $^{110}$Pd. (a) LDM potential
as a function of $\varepsilon$; (b) Components of the LDM potential, 
surface potential (above zero) and Coulomb potential (below zero), as 
a function of $\varepsilon$; (c) Di-nuclear shapes at $\varepsilon = 1$. 
The dashed, solid and dash-dotted lines are for $z = 1.4$, 1.67, 1.8, 
respectively. The slope $f$ equals to $0$ at $z = 1.67$.}
\label{fig-shape}
\end{center}
\end{figure}

The reasons why the potential as a function of $\varepsilon$ shows 
different features are displayed in 
Fig. \ref{fig-shape}. The potentials in this figure are referred to 
the potential at the spherical compound state. At small relative distances 
$z$ (the red dashed line), reducing $\varepsilon$ from $\varepsilon = 1$ 
means decreasing the surface and thus the corresponding surface potential, 
while the Coulomb potential becomes larger (Fig. \ref{fig-shape}(b)). 
Since the surface potential reduces faster than the increasing of the
Coulomb potential, the total potential becomes smaller with reducing 
$\varepsilon$, as shown by the red dashed line in 
Fig. \ref{fig-shape}(a). When the distance between the projectile and 
target increases (black solid line in Fig. \ref{fig-shape}) the decreasing 
speed of the surface area becomes smaller. At a critical distance $z_c$ 
(for instance $z_c = 1.67$ for $^{110}$Pd + $^{110}$Pd), the decreasing 
of the surface potential equals the increasing of the Coulomb potential
with reducing $\varepsilon$ around $\varepsilon$ = 1, so that the slope 
at $\varepsilon = 1$ becomes zero, as shown by the solid line in 
Fig. \ref{fig-shape}(a)(b). While at larger $z$, shown by the blue dash-dotted
line, the surface area is stable with reducing $\varepsilon$ around 
$\varepsilon$ = 1 and thus the total potential is increasing while neck 
starts to evolve from $\varepsilon=1$. 

\begin{figure}[ptbh]
\begin{center}
\includegraphics[height=2.9431in,width=3.8688in]{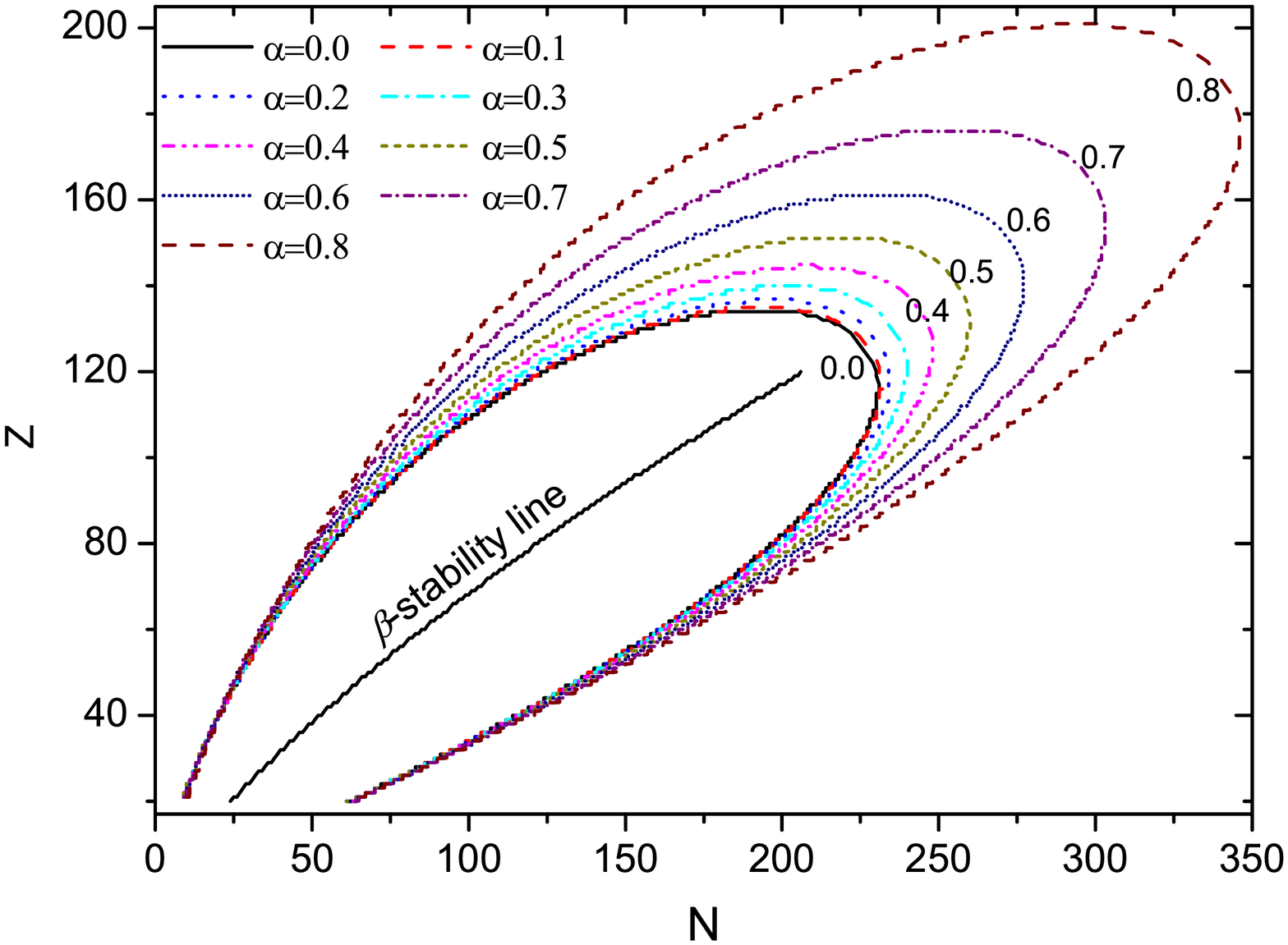}
\caption{The fusion reactions are classified as hindered (outside the lines)
or not-hindered (inside the lines) reactions in the neck evolution. On the
lines: $dV/d\varepsilon|_{(\varepsilon=1,z=z_{0})}$ = 0. The $Z$ and $N$ are
the proton number and neutron number of the compound nucleus, respectively.
See text for more details.}
\label{fig-2}
\end{center}
\end{figure}

In fact we only need to know the situation at contact ($z=z_{0}$). 
By simply checking the value of $f$ at $z=z_{0}$, we know whether the 
evolution of the neck degree of freedom is hindered or not. With this 
method, different combinations of the $A_{\rm{p}}$ and $A_{\rm{t}}$ 
for different neutron number $N(=N_{\rm{p}}+N_{\rm{t}})$ and proton
number $Z(=Z_{\rm{p}}+Z_{\rm{t}})$ of a compound nucleus are 
systematically studied and the results are shown in Fig. \ref{fig-2}.
The borderlines, corresponding to different $\alpha$, are plotted 
connecting the reactions for which $f$ equals 0 at $\varepsilon=1$ 
and $z=z_{0}$. The line of $\beta$-stability is also plotted. Inside
the borderlines, the slope $f$ is greater than 0 and the reactions 
are not hindered, while outside the lines it is the opposite.

Comparing the area encircled with different lines,
it also shows that the number of reactions without neck fusion hindrance
for small $\alpha$ is smaller than for large $\alpha$. This is due to the 
fact that more symmetric reactions correspond to larger Coulomb repulsion
force which leads the Coulomb potential to increase more quickly with
reducing $\varepsilon$. Since the encircled area is very large even for
$\alpha$ = 0, the neck evolution is not hindered for most of the realistic 
reactions.

\section{Fusion hindrance in the radial evolution}

\subsection{Average of the neck parameter}

We have just concluded that the neck evolution is not hindered for most 
of the realistic reactions. In fact, once at contact a strong potential 
slope drives the neck degree of freedom toward a mono-nucleus shape. On 
the other hand, the radial and mass-asymmetry degrees of freedom face a 
potential barrier for systems with hindered fusion. As a consequence, 
the neck degree of freedom evolves faster than the other two. It has 
reached equilibrium before the diffusion over the potential barrier has 
started \cite{abeneck,boilley1,boilley2}. As most of the di-nucleus evolve 
along $z$ and $\alpha$ direction with an equilibrated neck parameter, we 
study the fusion hindrance in the rest two degrees of freedom at the 
equilibrated value of $\varepsilon$. 

\begin{figure}[ptbh]
\begin{center}
\includegraphics[height=2.8624in,width=2.829in]{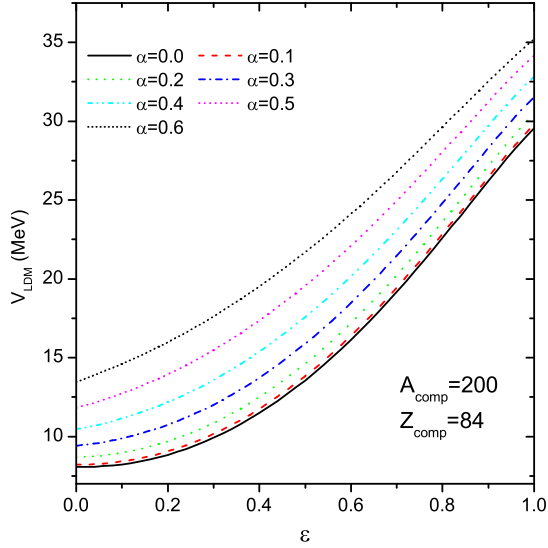}
\caption{LDM potential as a function of $\varepsilon$ for different 
symmetric or asymmetric reactions to form $^{200}$Po.}
\label{fig-aver-neck}
\end{center}
\end{figure}

The neck parameter at equilibrium can be determined through the average of
$\varepsilon$ via
\begin{equation}
\left\langle \varepsilon\right\rangle =
\frac{\int\varepsilon w(\varepsilon)d\varepsilon}
{\int w(\varepsilon)d\varepsilon},
\label{eq-3.5} 
\end{equation}
where $w(\varepsilon)=e^{-V(\varepsilon)/T}$, and $T$ is the temperature 
of the system. As shown in Fig. \ref{fig-aver-neck}, $\left\langle 
\varepsilon\right\rangle $ differs slightly for different reactions. 
In most cases, $\left\langle \varepsilon\right\rangle $ is close to 0.1.
Therefore we take $\varepsilon=0.1$ in the next calculations. This value 
is also used in the calculation of fusion cross section for $^{48}$Ca
induced reactions in the two-step model, and with which 
the residue cross sections are calculated combining the statistical 
evaporation model \cite{shen2}. By introducing a shell energy reduction 
factor, the calculated residue cross sections show good agreement with 
experimental data from Ca + U to Ca + Bk.

\subsection{Radial fusion hindrance}

Based on the LDM, when the projectile and target evolve from the contact 
point to the compound state, the parameters \{$\alpha$, $z$\} change from 
\{$\alpha_0$, $z_0$\} to \{$\alpha$ = 0, $z$ = 0\} in the same time. However
because the LDM potential as a function of $\alpha$ at fixed $z$ and 
$\varepsilon$ is similar to a parabolic potential centered at $\alpha=0$,
the evolving system is not hindered in the mass-asymmetry
direction. Therefore we only need to consider the fusion hindrance in 
the radial direction. Since the friction in the $\alpha$ and $z$ 
direction is close to each other, the radial direction cannot be 
de-coupled from the mass-asymmetry direction. Thus in the analysis
of the radial fusion hindrance we will consider $\alpha$ and $z$ 
direction in the same footing.

\begin{figure}[ptb]
\begin{center}
\includegraphics[height=2.9093in,width=2.8171in]{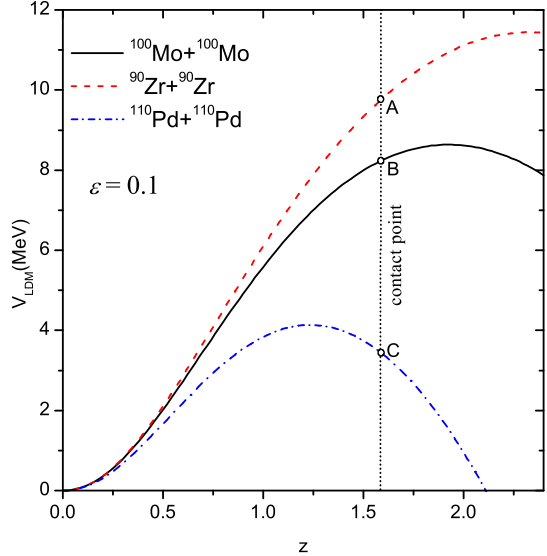}
\caption{The relation between the contact point and the saddle point 
for different reactions. The abscissa and ordinate stands for the 
distance between the two touching nuclei and LDM potential 
($\varepsilon = 0.1$), respectively. The vertical dashed line corresponds 
to the contact point $z = 1.587$ for symmetric reactions.}
\label{fig-rfh-sym}
\end{center}
\end{figure}

It is easier to understand the radial fusion hindrance in symmetric 
reactions by simply comparing the contact point and the saddle point, 
shown in Fig. \ref{fig-rfh-sym}. In this figure $\varepsilon$ has been 
fixed to 0.1. For reactions with light nuclei as $^{90}$Zr + $^{90}$Zr 
and $^{100}$Mo + $^{100}$Mo, the saddle points are located at the right 
hand side of the contact line ($z_{0}$= 1.587), which means that projectile 
and target at contact will evolve towards each other automatically by the 
driving force of the LDM potential and finally form the compound state. 
While for the reaction $^{110}$Pd + $^{110}$Pd, it is opposite: when the 
two touching $^{110}$Pd reach to the neck equilibrium (point ``C'' in 
Fig. \ref{fig-rfh-sym}), they need to overcome an extra-barrier to reach 
the compound state. In other words, the reaction is hindered, or the 
reaction has radial fusion hindrance. This is another fusion hindrance 
at the contact state besides the neck fusion hindrance. 

However this method cannot be applied to asymmetric reactions 
because the saddle point changes into a ridgeline. But similar 
to the symmetric case, as we can imagine, if the contact point 
is located inside (left hand side of) the ridge line the reaction 
will reach the compound state without radial fusion hindrance, 
and v.v. Unfortunately, it is not easy to get the ridgeline from 
a potential surface. To find the fusion path by diffusion, one 
has to solve a Langevin equation as shown in Eq. (6) of 
Ref. \cite{shen1}. Here, we employ another method: switching off 
the random force, setting the initial momentum in $z$ and $\alpha$ 
direction as zero and then studying the evolution track of the 
reaction by solving the Newtonian equation. The friction tensor is 
calculated with the wall-and-window formula \cite{blocki2}.
If the system reaches the compound state it means that the contact
point is located inside the ridgeline, and thus the system is not 
hindered. Contrarily the contact point is located outside the 
ridgeline and extra-push energy is necessary to form a compound 
nucleus. In fact this method can also be used for the symmetric case.

\begin{figure}[ptb]
\begin{center}
\includegraphics[height=2.8153in,width=3.8447in]{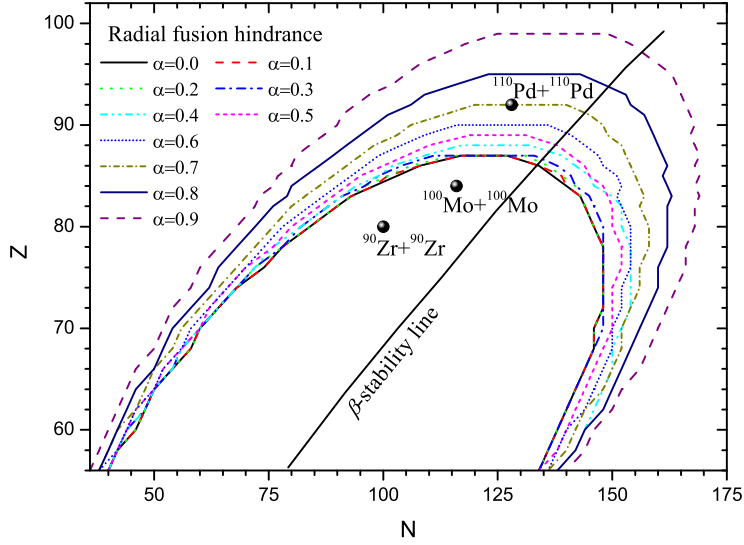}
\caption{Borderlines of the radial fusion hindrance. Reactions 
located inside the lines do not have radial fusion hindrance, 
while reactions located outside the lines are hindered in the 
radial direction. Three reactions and the line of $\beta$-stability 
are also plotted. The abscissa and ordinate stands for the neutron 
number $N$ and proton number $Z$ of the compound nucleus, respectively. }
\label{fig-rfh}
\end{center}
\end{figure}

With this method we study the radial fusion hindrance in symmetric 
and asymmetric reactions. Fig. \ref{fig-rfh} shows the location of 
the borderline between hindered and non-hindered reactions for 
various asymmetries $\alpha$. For the reactions located on the 
borderlines, the contact points are just located on the ridgeline, 
while for the reactions located inside the borderlines and outside 
the borderlines, the contact points are located inside and outside 
of the ridgeline, respectively. Therefore all of the reactions 
located outside of the ridgelines need extra-push energy to reach 
the compound state, or in other words, these reactions have radial 
fusion hindrance.

Fig. \ref{fig-rfh} also shows that the number of 
reactions without fusion hindrance at large $\alpha$ is larger than 
that at small $\alpha$ as it is well known experimentally. This can 
be easily understood: more symmetric reactions (with small $\alpha$) 
have larger repulsive Coulomb potential ($\propto1-\alpha^{2}$) and 
the ridgeline is closer to or even shifted into the left side of the 
contact points. In the same time, with reducing $\alpha$, the contact 
point $z_{0}$ becomes larger, as indicated in Eq. (\ref{eq-2.5}). 
This further reduces the number of reactions without fusion hindrance 
at more symmetric reactions.

It is commonly known from experiment that 
the reactions $^{90}$Zr + $^{90}$Zr, $^{100}$Mo + $^{100}$Mo are not 
hindered in the fusion reaction while the reaction $^{110}$Pd + $^{110}$Pd 
is \cite{keller,schmidt,morawek}. The three reactions are also plotted 
in Fig. \ref{fig-rfh}: Zr+Zr and Mo+Mo are located inside the line of 
$\alpha = 0$ while Pd+Pd is located outside, showing the agreement with 
theoretical analysis. It is interesting to see that the lower two dots 
($^{180}$Hg: Zr+Zr and $^{200}$Po: Mo+Mo) is located inside all of the
lines, showing that all of the reactions to form $^{180}$Hg 
and $^{200}$Po are not hindered. While the reactions to form $^{220}$U 
(the top dot marked by ``$^{110}$Pd+$^{110}$Pd'') is different: because
compound nucleus is located outside of the line with $\alpha <$ 0.7 and 
inside of the line with $\alpha >$ 0.7, then the reactions with 
$\alpha <$ 0.7 are hindered while the reactions with $\alpha >$ 0.7 
are not.

Fig. \ref{fig-rfh} also shows that the 
number of reactions without fusion hindrance is limited for each asymmetric
parameter $\alpha$, and each line for the corresponding $\alpha$ has a 
$Z_{\max}$. For example, $Z_{\max} = 87$ for $\alpha = 0.0$ and $Z_{\max} = 90$ 
for $\alpha = 0.6$. According to the extra-push theory\cite{swiatecki}, 
$Z_{\max} = 86$ for $\alpha = 0$ and $Z_{\max} = 110$ for $\alpha = 0.6$ 
if $x_{\rm{th}} = 0.7$. The comparison shows that the realistic 
calculations give smaller $Z_{\max}$ for larger $\alpha$, while for the 
reactions with smaller $\alpha$, the two models give similar $Z_{\max}$.

Comparing the Figs. \ref{fig-2} and \ref{fig-rfh}, it is also found that 
the encircled area in Fig. \ref{fig-rfh} is much smaller than the area in 
Fig. \ref{fig-2}, and the former one is completely included in the latter one. 
This indicates that the reactions without radial fusion hindrance are 
apparently without neck fusion hindrance, while the reactions with neck 
fusion hindrance have obviously radial fusion hindrance.

\begin{figure}[ptb]
\begin{center}
\includegraphics[height=2.0865in,width=2.8905in]{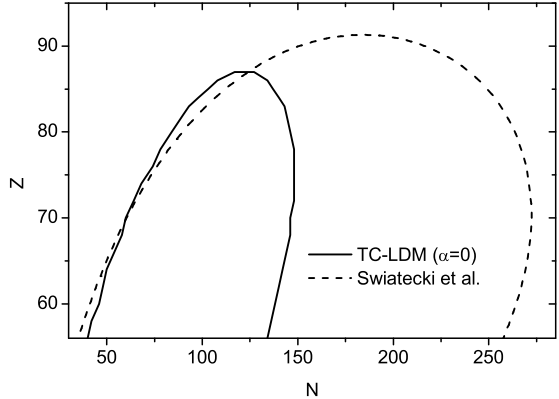}
\caption{Comparison of the radial fusion hindrance between the two-center 
LDM model (solid line) and the parameterized LDM model by Swiatecki et al. 
(dashed line) at $\alpha = 0$.}
\label{fig-compare}
\end{center}
\end{figure}

The symmetric case can be compared with the parameterized LDM proposed by 
Swiatecki et al.\cite{swiatecki}. In Ref. \cite{swiatecki}, the proton 
number $Z$ and the mass number $A$ of the compound nucleus, for which 
the contact point of the symmetric reaction is located on the saddle 
point of the parameterized LDM, satisfies
\begin{equation}
\frac{676.48-38.534\sqrt{172.88-A/2}}{5+1352.94/A}
\leq Z\leq\frac{676.48+38.534\sqrt{172.88-A/2}}{5+1352.94/A}.
\label{eq-4}
\end{equation}
Plotting together the above Eq. (\ref{eq-4}) and the borderline
at $\alpha = 0$ of Fig. \ref{fig-rfh}, we get Fig. \ref{fig-compare}. 
It is clear that the lower limit of $N$ for fixed $Z$ is almost the 
same between two models, but the upper limit of $N$ is quite different: 
the parameterized LDM gives much larger $N_{\max}$ than the two-center 
LDM model. This is probably caused by the too few parameters used to describe 
the so complicated liquid drop energy in Ref. \cite{swiatecki}.

\section{Conclusion}

Fusion reactions between two colliding heavy ions may be hindered or not. 
This is explained by the appearance of an inner barrier after the contact 
point that hampers the formation of a compound nucleus. The liquid drop 
potential landscape described with the two-center parameterization for 
the deformation allows us to study the frontier between hindered and 
non-hindered reactions as shown in Fig. \ref{fig-2} and \ref{fig-rfh}. 
The neck and radial degrees of freedom may both have to overcome an 
inner barrier to reach the compound shape. For each of them, we can draw a 
borderline that classifies reactions with respect to the hindrance.

It appears that for available nuclei, the neck degree of freedom is not 
hindered. A strong potential slope always drives it toward a mono-nucleus 
shape. It is not the case for the radial degree of freedom. We obtained a 
borderline that is far more restrictive, i.e. the hindrance to the fusion 
appears for compound nuclei having a charge larger than $Z=87$ when they 
are formed by a symmetric reaction. The frontier is located at larger 
values of $Z$ for asymmetric reactions, as it is observed experimentally.

The exact location of the border between hindered and non-hindered reactions
might depend on the parameterization. It would be therefore very useful 
to have some experimental results to assess the models.

Finally, it is important to note that in this paper, 
we have just studied the appearance of the hindrance to the fusion. 
The amplitude of the hindrance, i.e. the extra-energy that is necessary 
to the system to fuse, also depends on dynamical parameters as the 
friction coefficient \cite{abe3,boilley0}. 

\begin{acknowledgments}

The work is supported in part by the National Natural Science Foundation of
China (Nos. 10905021, 10979023, 10979024), the Key Project of the Ministry of
Education of China (No. 209053), the Zhejiang Provincial Natural Science
Foundation of China (No. Y6090210), and the Qian-Jiang Talents Project of
Zhejiang Province (No. 2010R10102). The authors also acknowledge supports and
hospitality by RCNP Osaka University, GANIL, and Huzhou Teachers College,
which enable us to continue the collaboration. This work is also partly
supported by the C3S2 computing center in Huzhou Teachers College.

\end{acknowledgments}

\end{document}